\documentclass{aastex63}
\usepackage{hyperref}

\begin{document}

% General purpose macros
\newcommand{\lbol}{\mbox{$L_{bol}$}} % bolometric luminosity
\newcommand{\lsmm}{\mbox{$L_{smm}$}} % luminosity longward of 350 mic.
\newcommand{\lint}{\mbox{$L_{int}$}} % internal luminosity
\newcommand{\lacc}{\mbox{$L_{acc}$}} % accretion luminosity
\newcommand{\tbol}{\mbox{$T_{bol}$}} % bolometric temperature
\newcommand{\lbolsmm}{\mbox{$L_{bol}/L_{smm}$}} % bol-to-smm luminosity
\newcommand{\lsmmbol}{\mbox{$L_{smm}/L_{bol}$}} % smm-to-bol luminosity
\newcommand{\av}{\mbox{$A_V$}} % Visual Extinction

% Macros for units
\newcommand{\degree}{\mbox{$^{\circ}$}}
\newcommand{\am}{\mbox{\arcmin}}
\newcommand{\as}{\mbox{\arcsec}}
\newcommand{\kms}{\mbox{kms$^{-1}$}}% km/s
\newcommand{\jybeam}{\mbox{Jy beam$^{-1}$}}% Jy/beam
\newcommand{\mjybeam}{\mbox{mJy beam$^{-1}$}}% Jy/beam
\newcommand{\um}{$\mu$m}
\newcommand{\lsun}{\mbox{L$_\odot$}}% Lsun
\newcommand{\msun}{\mbox{M$_\odot$}}% Msun
\newcommand{\rsun}{\mbox{M$_\odot$}}% Msun

% Footnote command
\makeatletter
\newcommand\footnoteref[1]{\protected@xdef\@thefnmark{\ref{#1}}\@footnotemark}
\makeatother

\title{Detection of a Disk Surrounding the Variably Accreting Young Star HBC722}

\author{Xi Yek (Zach)}
\altaffiliation{xyek@fredonia.edu}
\affiliation{Department of Physics, State University of New York at Fredonia, 280 Central Ave, Fredonia, NY 14063, USA}

\author{Michael M.~Dunham}
\altaffiliation{michael.dunham@fredonia.edu}
\affiliation{Department of Physics, State University of New York at Fredonia, 280 Central Ave, Fredonia, NY 14063, USA}
\affiliation{Center for Astrophysics $|$ Harvard~\&~Smithsonian, 60 Garden Street, Cambridge, MA 02138, USA}

\author{H\'ector G.~Arce}
\affiliation{Department of Astronomy, Yale University, P.O. Box 208101, New Haven, CT 06520-8101, USA}

\author{Tyler L.~Bourke}
\affiliation{SKA Organisation, Jodrell Bank, Lower Withington, Macclesfield, Cheshire SK11 9FT, UK}

\author{Xuepeng Chen}
\affiliation{Purple Mountain Observatory, Chinese Academy of Sciences, 2 West Beijing Road, Nanjing 210008, China}

\author{Joel D.~Green}
\affiliation{Space Telescope Science Institute, 3700 San Martin Dr., Baltimore, MD 02138, USA}

\author{\'Agnes K\'osp\'al}
\affiliation{Konkoly Observatory, Research Centre for Astronomy and Earth Sciences, Konkoly-Thege Mikl\'os \'ut 15-17, 1121 Budapest, Hungary}
\affiliation{Max Planck Institute for Astronomy, K\"onigstuhl 17, 69117 Heidelberg, Germany}
\affiliation{ELTE E\"otv\"os Lor\'and University, Institute of Physics, P\'azm\'any P\'eter s\'et\'any 1/A, 1117 Budapest, Hungary}

\author{Steven N.~Longmore}
\affiliation{Astrophysics Research Institute, Liverpool John Moores University, 146 Brownlow Hill, Liverpool L3 5RF, UK}

\submitjournal{Research Notes of the AAS}
%\date{May 2020}

\shorttitle{ALMA HBC722 Detection}
\shortauthors{Yek \& Dunham}

%\begin{document}
%\maketitle

\begin{abstract}
    We present new ALMA 233~GHz continuum observations of the FU Orionis Object HBC722.  With these data we detect HBC722 at millimeter wavelengths for the first time, use this detection to calculate a circumstellar disk mass of 0.024~\msun, and discuss implications for the burst triggering mechanism.
\end{abstract}

\section{Introduction}\label{sec_intro}
FU Orionis objects (hereafter FUors) are a group of pre-main-sequence stars that abruptly increase in brightness by several magnitudes, lasting decades or longer. There are about $10-20$ FUors confirmed by direct observation of their bursts, along with an approximately equal number of candidates found to display similar spectral characteristics \citep[e.g.,][]{reipurthaspin2010:fuors}. The FUors’ large amplitude bursts are credited to enhanced mass accretion from the surrounding circumstellar disk, with various triggering mechanisms being proposed over the years, e.g. gravitational and/or magnetorotational instabilities, thermal instabilities, or interactions with binary companions \citep[e.g.,][]{hartmann1985:fuors,hartmann1996:fuors,audard2014:ppvi}.  FUors may represent the late stages of a cycle of episodic accretion bursts and luminosity flares during the embedded phase \citep{dunham2014:ppvi}.  Thus, studying and characterizing each FUor is important for understanding how they fit into the general star formation process.

In this paper, we present new 1~mm continuum observations of the FUor HBC722 obtained with the Atacama Large Millimeter/Submillimeter Array (ALMA).  HBC722, located in the North American/Pelican Nebula Complex \citep[$d \sim 800$~pc;][]{zucker2020:distances}, was undetected in previous 1~mm continuum data from the Submillimeter Array (SMA), setting an upper limit for the disk mass of 0.05~\msun\ \citep[after rescaling to the current distance;][]{dunham2012:hbc722}. With our new ALMA data we present the first millimeter continuum detection of this object, and use our data to calculate the mass of its circumstellar disk.

\section{Observations \& Results}\label{sec_obs}
We obtained ALMA Cycle 2, Band 6, 12-m array observations of HBC722 on 2014 May 01 and 2015 May 03, with 34~--~36 operational antennas.  The array configuration provided projected baseline lengths of 12$-$450~m and a synthesized beam size (assuming natural weighting) of 1.67$"$~$\times$~0.89$"$ at a position angle of 5.82\degree\ (measured east of north).  %No Atacama Compact Array (ACA) 7-m or total-power observations were obtained in this program.
To measure the continuum with the widest possible bandwidth while avoiding contamination from bright lines, the four spectral windows were centered at 224, 226, 240, and 242~GHz, and were configured to provide 128 channels over 1.875~GHz bandwidth.
%, corresponding to a channel spacing of 14.6~MHz (18.9~\kms\ at 233~GHz).
The resulting continuum image is centered at 233~GHz (1.29~mm) and has a total bandwidth of 7.5~GHz.

Calibration was performed using the Common Astronomy Software Applications (CASA) package\footnote{Available at \href{http://casa.nrao.edu}{http://casa.nrao.edu}}, following the standard techniques described in \citet{petry2014:alma} and \citet{schnee2014:alma}.  We then applied two rounds of phase self-calibration, followed by one round of amplitude self-calibration.  The effective solutions intervals used for self-calibration were 390~s (first round of phase calibration), 30~s (second round of phase calibration), and 900~s (amplitude calibration).  After applying the amplitude self-calibration we verified that the noise in the image decreased while the total fluxes of the detected objects did not change.  The self-calibrated image has a 1$\sigma$ rms of 0.057~mJy~beam$^{-1}$.

%Our self-calibration decreased the 1$\sigma$ rms of the continuum image by a factor of 2.1, and it improved the peak signal-to-noise ratio of the HBC722 detection (described below) by a factor of six.

%The first round of phase self-calibration used an infinite solution interval so that all consecutive integrations on HBC722 before looping back to the gain calibrator are averaged together.  For our observations, this corresponds to an effective solution interval of 6.5~minutes (390~s).  The second round of phase self-calibration used a 30~s solution interval, and the single round of amplitude self-calibration used a 900~s solution interval.  

\begin{figure}[t]
    \centering
    \includegraphics[width=12cm]{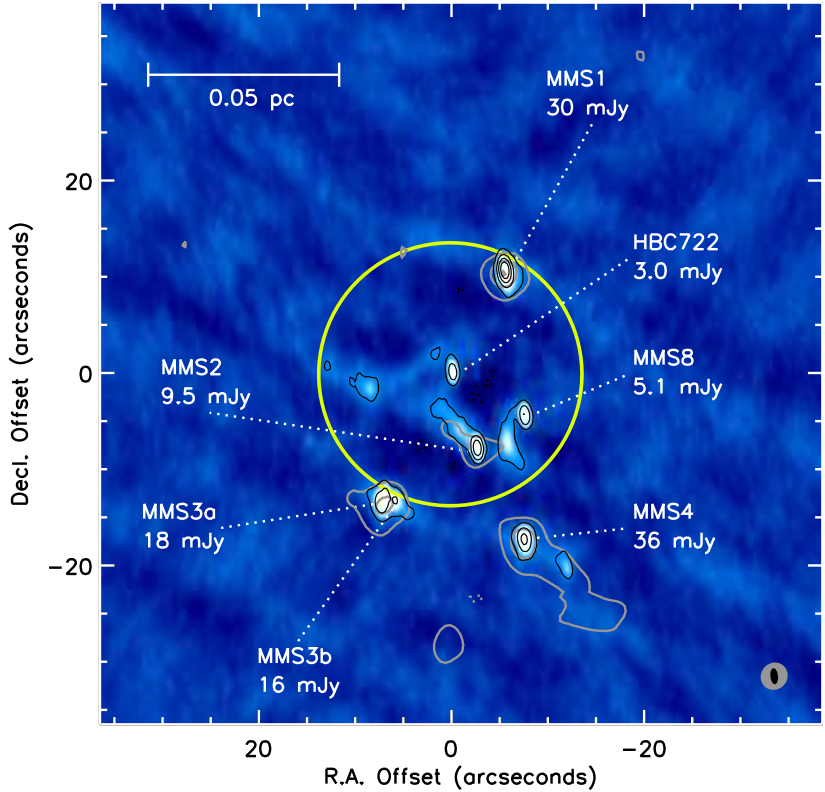}
    \caption{ALMA 233~GHz continuum image of HBC722.  The yellow circle shows the ALMA primary beam.  The black contours show the ALMA continuum data at levels of 5, 20, 50, and 100 times the 1$\sigma$ rms noise, whereas the gray contours show the SMA continuum data at approximately the same frequency from \citet{dunham2012:hbc722}, starting at 3$\sigma$ and increasing in steps of 5$\sigma$.  The ALMA and SMA synthesized beams are shown as the black and gray ellipses, respectively.  Detections from \citet{dunham2012:hbc722} are labeled, with annotations reporting their integrated flux densities from the primary beam corrected ALMA data.  MMS3 from \citet{dunham2012:hbc722} is resolved as two sources in our higher-resolution ALMA data, MMS5, MMS6, and MMS7 from \citet{dunham2012:hbc722} are too far outside the ALMA primary beam to be robustly detected, and MMS8 is a new ALMA detection that was only marginally detected in the SMA data.}
    \label{fig_continuum}
\end{figure}

In Figure~1, which displays the self-calibrated ALMA 233~GHz continuum image, HBC722 is detected with a peak signal-to-noise ratio of 51.  With a peak intensity of 2.9~mJy~beam$^{-1}$, the ALMA detection of HBC722 is consistent with the previous SMA non-detection from \citet{dunham2012:hbc722} at approximately the same frequency.

\section{Analysis \& Conclusions}\label{sec_analysis}
We used the CASA tool {\sc gaussfit} to fit an elliptical Gaussian in the image plane to all of the detected sources in the ALMA continuum image.  The resulting integrated flux densities are included as annotations in Figure 1.  For HBC722, we then calculated the total circumstellar disk mass $M$ as: 
\begin{equation}
\label{eqn:mass equation}
M=100 \frac{d^2 S_\nu}{B_\nu (T_D) \kappa_\nu \quad},    
\end{equation}
where $d=800$~pc, $S_\nu$ is the measured integrated flux density, $B_\nu (T_D)$ is the Planck function at the isothermal dust temperature $T_D$, $\kappa_\nu$ is the opacity of the dust, and the factor of 100 is the assumed gas-to-dust ratio. We adopt the dust opacities of \citet{ossenkopf1994:oh5} for thin ice mantles after 10$^5$~yrs of coagulation at a gas density of 10$^6$~cm$^{-3}$, giving $\kappa_\nu=0.901$~cm$^2$~g$^{-1}$ at 233~GHz. Assuming $T_D=30$~K, Equation \ref{eqn:mass equation} gives a total circumstellar disk mass for HBC722 of 0.024~\msun.  With typical uncertainties of a factor of a few \citep{dunham2014:disks}, this result is in good agreement with the total gas mass of 0.03~\msun\ found by \citet{kospal2016:hbc722}.

With a burst accretion rate of 10$^{-6}$~\msun~yr$^{-1}$ \citep{kospal2011:hbc722}, it would take 24,000~yr to drain a 0.024~\msun\ disk.  Thus this disk mass is sufficient to power the current accretion burst.  With a known stellar mass of approximately 0.5~\msun\ \citep{cohen1979:hbc722}, our calculated disk mass implies that HBC722 has a disk-to-star mass ratio of approximately 5\%.  While this is a factor of ten higher than the median disk-to-star mass ratio for T Tauri stars \citep{andrews2005:disks}, it is still marginally too low for gravitational instabilities to serve as the burst triggering mechanism, as such instabilities likely require disk-to-star mass ratios of 10\% or higher.  However, given the uncertainties in the calculated mass, and the possibility that the ALMA continuum detection is optically thick, gravitational instabilities remain a possible triggering mechanism for this object.  
Future analysis using radiative transfer modeling is required to better determine the true mass of the HBC722 disk.

\acknowledgments
This paper makes use of the following ALMA data: ADS/JAO.ALMA\#2013.1.00088.S. ALMA is a partnership of ESO (representing its member states), NSF (USA) and NINS (Japan), together with NRC (Canada), MOST and ASIAA (Taiwan), and KASI (Republic of Korea), in cooperation with the Republic of Chile. The Joint ALMA Observatory is operated by ESO, AUI/NRAO and NAOJ.  The National Radio Astronomy Observatory is a facility of the National Science Foundation operated under cooperative agreement by Associated Universities, Inc.  This project has received funding from the European Research Council (ERC) under the European Union's Horizon 2020 research and innovation programme under grant agreement No 716155 (SACCRED).  This research has made use of NASA's Astrophysics Data System (ADS) Abstract Service and the IDL Astronomy Library hosted by the NASA Goddard Space Flight Center.

\bibliographystyle{aasjournal.bst}
\bibliography{citations}

\end{document}